\documentclass[10pt,letterpaper]{article}
\usepackage[top=0.85in,left=2.75in,footskip=0.75in]{geometry}

\usepackage{amsmath,amssymb}

\usepackage{changepage}

\usepackage[utf8x]{inputenc}

\usepackage{textcomp,marvosym}

\usepackage{cite}

\usepackage{nameref,hyperref}

\usepackage[right]{lineno}

\usepackage{microtype}
\DisableLigatures[f]{encoding = *, family = * }

\usepackage[table]{xcolor}

\usepackage{array}

\newcolumntype{+}{!{\vrule width 2pt}}

\newlength\savedwidth

\raggedright
\usepackage[aboveskip=1pt,labelfont=bf,labelsep=period,justification=raggedright,singlelinecheck=off]{caption}

\makeatletter
\renewcommand{\@biblabel}[1]{\quad#1.}
\makeatother

\usepackage{lastpage,fancyhdr,graphicx}
\fancyheadoffset[L]{2.25in}
\fancyfootoffset[L]{2.25in}
\begin{document}
\vspace*{0.2in}

\begin{flushleft}
{\Large
\textbf\newline{Multi-stable oscillations in cortical networks with two classes of inhibition} 
}
\newline
\\
Arnab Dey Sarkar\textsuperscript{1,*},Bard Ermentrout\textsuperscript{1}

\bigskip
\textbf{1}Department of Mathematics,
University of Pittsburgh,Pittsburgh PA 15260, USA

\bigskip
* ARD129@pitt.edu

\end{flushleft}
\section*{Abstract}
In the classic view of cortical rhythms, the interaction between excitatory pyramidal neurons (E) and inhibitory parvalbumin neurons (I) has been shown to be sufficient to generate gamma and beta band rhythms.  However, it is now clear that there are multiple inhibitory interneuron subtypes and that they play important roles in the generation of these rhythms. In this paper we develop a spiking network that consists of populations of E, I  and an additional interneuron type, the somatostatin (S) internerons that receive excitation from the E cells and inhibit both the E cells and the I cells.   These S cells are modulated by a third inhibitory subtype, VIP neurons that receive inputs from other cortical areas.  We reduce the spiking network to a system of nine differential equations that characterize the mean voltage, firing rate, and synaptic conductance for each population and using this we find many instances of multiple rhythms within the network. Using tools from nonlinear dynamics, we explore the roles of each of the two classes of inhibition as well as the role of the VIP modulation on the properties of these rhythms.

\section*{Author summary}
Rhythmic dynamics in the cortex are crucial for information processing, sensory integration, and cognition. In this paper, we look at a model  network consisting of a population of excitatory neurons and two distinct populations of inhibitory neurons. We show that the interactions between these three populations gives rise to multiple coexistent rhythms. We also present a greatly simplified model that can be tuned to have similar properties.  Our computational model may provide a mechanism for  the experimental appearance of multiple rhythms in the same cortical circuit.


\section*{Introduction}

In recent years, substantial research has focused on the dynamics of \textbf{E+I neuron networks}, comprising excitatory pyramidal neurons (E) and inhibitory parvalbumin-positive interneurons (PV-INs, I), which are known to generate oscillatory patterns crucial for various neural functions  \cite{GX2014Gamma, yi, mi}. These studies, particularly those centered on gamma ($\gamma$) oscillations, have elucidated how the balance of excitation and inhibition shapes neural network activity. However, an increasing body of evidence highlights the importance of other inhibitory interneuron populations, particularly \textbf{somatostatin-expressing interneurons (SST-INs)}, which exhibit distinct temporal properties and connectivity patterns compared to PV-INs \cite{chen2017somatostatin, yc, onorato2025gamma}.

SST-INs play a unique role in cortical circuits by providing dendritic inhibition, thereby influencing input integration and network excitability differently than perisomatic inhibition by PV-INs. Their involvement has been implicated in modulating oscillations across various frequency bands, including beta ($\beta$) rhythms, which are crucial for motor control, working memory, and attention. Understanding the interaction between excitatory neurons, PV-INs, and SST-INs is critical for a more comprehensive view of cortical dynamics, especially in contexts where multiple oscillatory regimes coexist and interact \cite{chen2017somatostatin,br,ABV}.

Using computational modeling and in vitro electrophysiological recordings from cortical slices ,\cite{pse} studies have shown that beta1 and beta2 rhythms arise from overlapping but distinct network dynamics, with excitatory pyramidal neurons primarily supporting beta1, PV-INs enhancing beta2 through strong inhibitory drive, and SST-INs modulating the phase coupling between the two bands by precisely regulating distal dendritic inhibition and shifting the excitation-inhibition balance to promote cross-frequency coherence \cite{rkc, gk,racine2021somatostatin,kopell2011neuronal,unknown2019parietal}. Motivated by these biological observations, we sought to investigate the emergence of multirhythmicity in cortical networks using a mathematical model incorporating two distinct types of inhibition—PV and somatostatin (SST)—with vasoactive intestinal peptide (VIP) interneurons acting as a modulatory input that selectively inhibits SST cells.

Recent theoretical work has begun to address how multiple inhibitory cell classes interact to control cortical rhythms and network stability. In particular, Edwards et al. \cite{edwards24} constructed a biophysically inspired spiking network model incorporating three distinct neuronal populations: excitatory pyramidal neurons (E), parvalbumin-positive (PV) fast-spiking interneurons, and somatostatin-positive (SST) interneurons. Their study was motivated by recent in vivo findings in the primary visual cortex (V1), which suggest that SST and PV interneurons have distinct temporal firing patterns and phase-locking behavior during gamma oscillations \cite{onorato2025gamma, chen2017somatostatin}.
Through a series of numerical simulations and phase-locking analyses, they revealed that PV and SST interneurons exert qualitatively distinct effects on network dynamics. PV-INs strongly synchronize excitatory neurons via perisomatic inhibition, facilitating high-frequency gamma rhythms (40–80 Hz), while SST-INs contribute to slower inhibitory feedback loops that modulate gain and network stability. Crucially, their work demonstrated that alterations in SST-PV-E connectivity can give rise to rich dynamical regimes, including asynchronous irregular states, bistability, and multirhythmic oscillations. These findings align with earlier computational predictions \cite{kopell2011neuronal, gk, rkc}, and provide new insight into how SST-INs may gate transitions between different oscillatory modes.

Recent work by Tahvili et al. \cite{FMM} further advanced this line of research with the CAMINOS model, which dissected the causal contributions of PV and SOM interneurons. Their study demonstrated that PV cells are critical for controlling oscillation frequency and maintaining stability, while SOM cells regulate amplitude and promote slower rhythms. Importantly, they showed that balanced ratios of PV and SOM neurons yield the most stable cortical dynamics, whereas dominance of either class can destabilize the network, sometimes leading to pathological activity. This highlights how interneuron diversity not only enables rhythm generation but also safeguards against instability. Complementary work \cite{terwal2021} systematically analyzed all possible E–PV–SOM circuit motifs, identifying a taxonomy of oscillatory states—including theta-nested gamma, stable beta, and theta-locked bursting—that emerge only in networks with multiple interneuron subtypes. Their motif-level classification provides a broad map of circuit behaviors that complements our mechanistic focus on multistability and switching.

Inspired by this framework, we extend their approach by implementing a reduced mean-field model that captures the macroscopic dynamics of such three-population networks. Building on \cite{montbrio}, we incorporate two classes of inhibitory neurons—PV and SST—and investigate their role in shaping network rhythms across a range of input conditions and connectivity motifs. In contrast to prior two-population models \cite{wc72, be79, borgers03}, our formulation allows for a deeper analysis of the interactions between multiple inhibitory motifs and their impact on spectral composition, phase locking, and cross-frequency coupling.

The primary objective of this study is to explore the dynamics of a three-population neuronal network consisting of \textbf{excitatory neurons (E)}, \textbf{parvalbumin interneurons (I)}, and \textbf{somatostatin interneurons (S)} by constructing and analyzing both \textbf{population-level models} and \textbf{spiking network models}. We aim to investigate how the interplay between these three populations influences network oscillations and bifurcation structures, which may provide insight into cross-frequency interactions and bistability observed in experimental studies.

In our approach, we develop a \textbf{9-dimensional (9D) population model} to describe the evolution of excitatory, inhibitory, and somatostatin neural activity. We conduct a bifurcation analysis of the 9D system to characterize the changes in oscillatory patterns and the emergence of bistability in response to varying external inputs and connectivity strengths. Using this analysis, we identify parameter regimes where multiple stable oscillatory states coexist, which motivates the transition to a more detailed \textbf{spiking network model} using the theta neuron model framework.

The spiking model consists of 400 neurons in each population and incorporates heterogeneity through a theta-neuron transformation, allowing us to link the macroscopic dynamics of the population model to microscopic spiking activity. By simulating this model under varying external stimuli and synaptic coupling parameters, we demonstrate how oscillatory dynamics shift between large and small limit cycles, reflecting the bistable behavior observed in the population-level bifurcation analysis.

In summary, this paper presents a systematic exploration of the interaction between excitatory, inhibitory, and somatostatin populations in neural networks. Through a combination of mathematical modeling, bifurcation analysis, and spiking network simulations, we provide a deeper understanding of how the balance of excitation, inhibition, and dendritic modulation governs the complex oscillatory behavior of cortical circuits. Our results may have implications for understanding neural mechanisms underlying cognitive processes and disorders associated with altered beta and gamma oscillations.

\section*{Results}

\subsection*{Spiking model}

\begin{figure}
  \includegraphics[width=.9\textwidth]{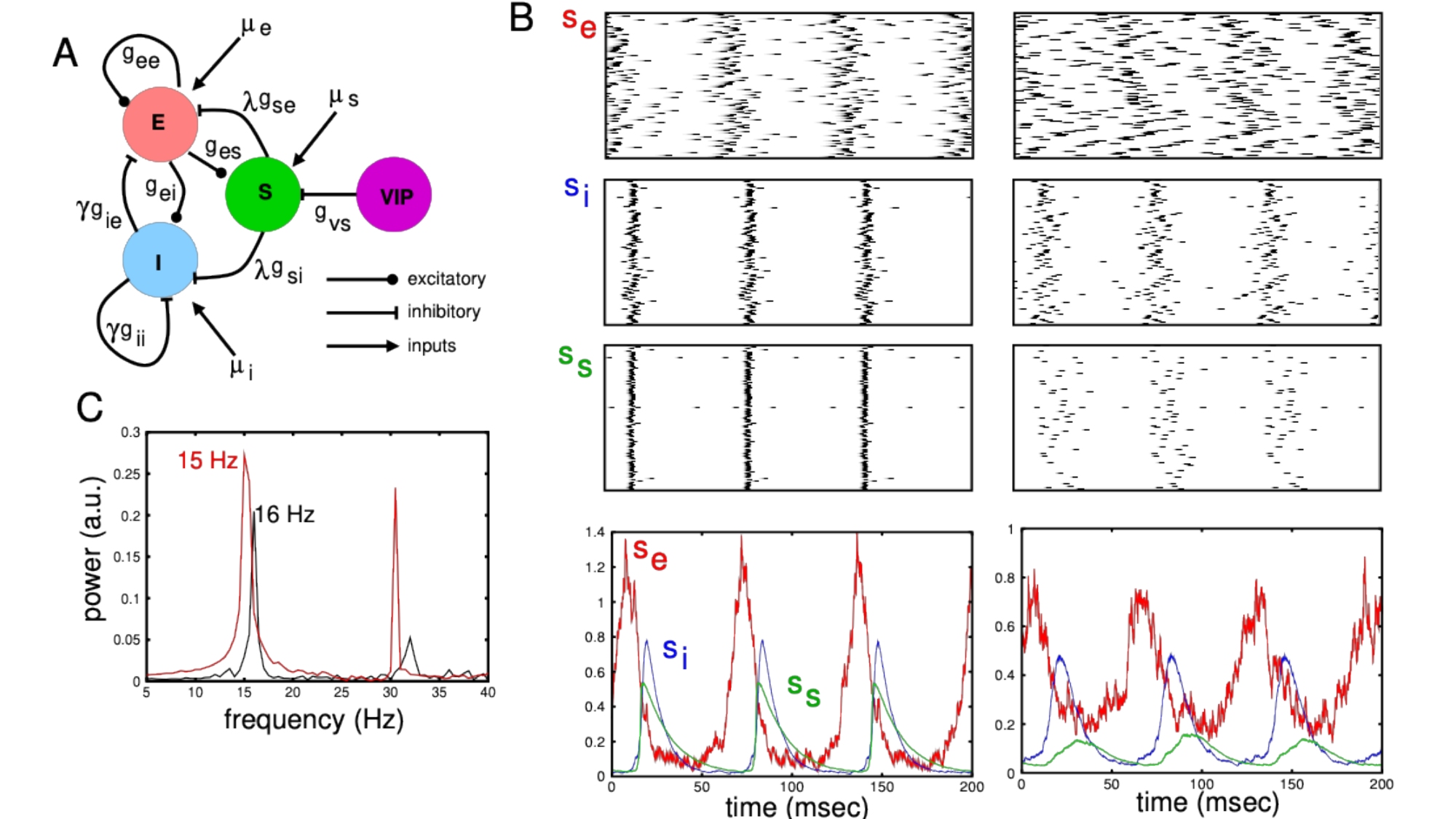}
  \caption{A. Model circuit showing connectivity between excitatory pyramidal cells (E), inhibitory parvalbumin cells (I), inhibitory somatostatin cells (S), and VIP inhibitory cells that modulate the excitability of the S cells. B. Behavior of the network for default parameters except $\mu_e=1.25$ showing two distinct rhythms in the same network. Top rows show raster plots of the three populations for the two rhythms and bottom shows the population synaptic response (filtered firing rates). (C) Power spectrum for the two rhythms showing peaks at 15 and 16 Hz.}
  \label{fig:1}
\end{figure}

We have created  network with excitatory cells (E) and three types of inhibition (I,S,VIP) as shown in Fig. \ref{fig:1}A \cite{veit23,bos20,edwards24}.  The VIP is represented as constant negative bias modulating the  S cells. Each E,I,S population is modeled by 400  all-to-all connected quadratic integrate-and-fire neurons (QIF) with connections mediated by exponentially decaying synapses (see Methods).  Each cell in the network receives a constant bias current, common among the neurons within the population (e.g. $\mu_e$) as well as a constant small random input taken from a Cauchy distribution.  The main difference between the population of S cells and the population of I cells is that the S synapses decay more slowly \cite{edwards24} and their connectivity pattern is different. (See Table \ref{tab:par} for a complete list of default parameters.)  Since the VIP input to the S cells is modeled here as being strictly modulatory, in the simulations and analysis, we incorporate it into $\mu_s$, the input into the S cells.

In Fig. \ref{fig:1}B, we drive the excitatory population with a tonic input and find, depending on how the network is initialized, two distinct rhythms which are close in frequency (15 vs 16 Hz) but differ a great deal in their coherence and amplitude.  In the left column, we show rasters of the the three populations as well as the population synaptic response ($s_{e,i,s}$) over a 200 millisecond period.  The excitatory population is dispersed but there is a clear rhythm (top and bottom left). Both inhibitory populations show very tight coherence shown clearly in both the rasters and the population responses.  A second rhythm can be found by starting the network at a different state.  This slightly higher frequency rhythm is smaller in amplitude and, in particular, the S population fires very weakly with a large spread.  Panel C shows the power spectrum of $s_e(t)$ taken from 2 seconds of simulation.  The lower frequency rhythm (left one in panel B) shows higher power for the principle as well as the higher harmonics. Indeed, after the second harmonic, the peak disappears for the lower power rhythm but remains over many harmonics for the higher power rhythm (not shown).  We want to emphasize two points: (i) these rhythms do not appear concurrently, rather they are both stable attractors; and (ii) the parameters for both rhythms are the same; this is a multistable system.

\begin{figure}
  \includegraphics[width=.8\textwidth]{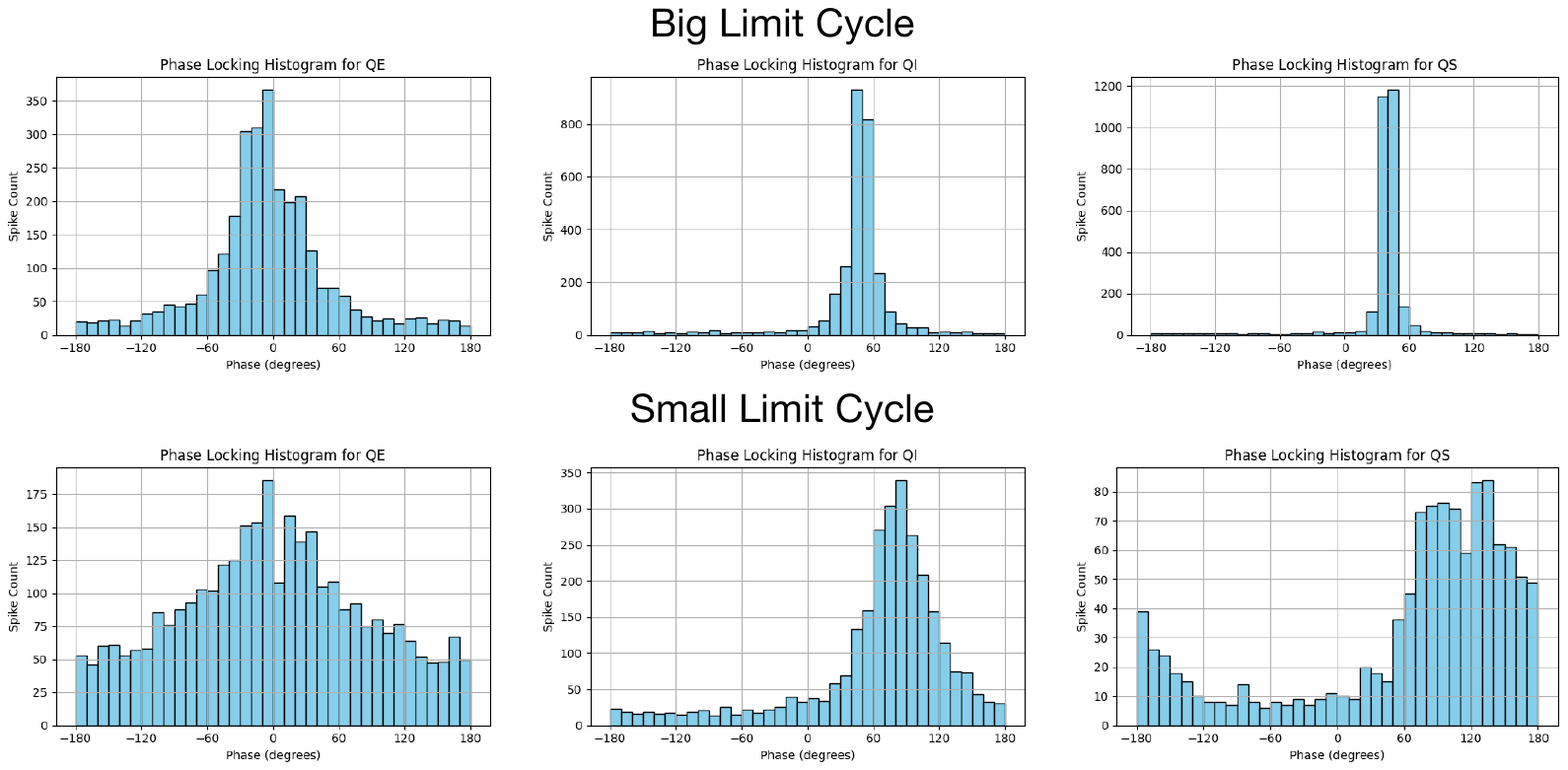}
  \caption{Spike phase relationships to the surrogate local field potential, $s_e(t)$ for E, I, and S populations for large (Top) and small (Bottom) limit cycles.}
  \label{fig:1b}
\end{figure}

In Fig. \ref{fig:1b}, we illustrate the relationships between the spike times of the three populations with $s_e(t)$ which we take as a surrogate for the local field potential. As has been found in other experimental and theoretial papers, \cite{chen2017somatostatin,FMM}, the two inhibitory classes have different phase relationships.  The S (SOM) population typically spikes later in phase than the I (PV) population. This is particularly evident in the ``small'' oscillation.  There are major differences between the large and small oscillations with respect to the number of spikes and how synchronoous they are.  As expected in the smaller amplitude rhythms, there are fewer spikes in total for all three populations.  More interestingly, the locking is much more spread out for the SOM cells in the smaller rhythm. In the large rhythm both S and I cells are phase-delayed by about $60^\circ$ and both are very sharply peakes. In contrast in the smaller amplitude rhythm the I cells peak a little later but the S cells peak substantially later in the cycle.  In both populations, they are also spread out more.

\subsection*{Mean field models of neural populations.}
To uncover the mechanisms for this multistability, the conditions which underly it and discover other behaviors of the network, we construct an exact mean-field model in the limit as the number of neurons in each population goes to infinity.  
Montbrio et al \cite{montbrio} developed a powerful method for reducing populations of quadratic integrate-and-fire (QIF) neurons to low-dimensional models which involve only differential equations for the mean firing rate, the mean potential, and the synaptic kinetics.  Their work was based on an equivalence between the QIF and the theta-model \cite{ke86} and an application of the Ott-Antonsen ansatz \cite{oa}.  Since these important papers, there have been many subsequent papers extending their approach to other types of neuronal networks.  In the Methods, we sketch the derivation of the mean field equations which we rewrite here:

\begin{eqnarray}
    \tau_{m,e} \dot{a_e} &=& 2 a_eb_e + \Delta_e \nonumber \\
    \tau_{m,e} \dot{b_e} &=& b_e^2-a_e^2 + \mu_e + g_{ee}s_e - \gamma g_{ie}s_i-\lambda g_{se}s_s \nonumber \\
    \tau_{e} \dot{s_e} &=& -s_e+a_e/\pi \nonumber \\
    \tau_{m,i} \dot{a_i} &=& 2 a_ib_i + \Delta_i \nonumber \\
    \tau_{m,i} \dot{b_i} &=& b_i^2-a_i^2 + \mu_i + g_{ei}s_e - \gamma g_{ii}s_i-\lambda g_{si}s_s  \\
    \label{eq:eisoa}
    \tau_{i} \dot{s_i} &=& -s_i+a_i/\pi. \nonumber \\
    \tau_{m,s} \dot{a_s} &=& 2 a_sb_s + \Delta_s \nonumber \\
    \tau_{m,s} \dot{b_s} &=& b_s^2-a_s^2 + \mu_s + g_{es}s_e  \nonumber \\
    \tau_{s} \dot{s_s} &=& -s_s+a_s/\pi. \nonumber
\end{eqnarray}
For each population, the variable $a$ is proportional to the mean firing rate, $b$ is mean voltage, and $s$ is the synaptic output of each population.   
We have included two parameters, $\gamma$ and $\lambda$ which will allow us to vary the influence of the respective I and S inhibitory populations.  As noted above, the influence of the VIP population is incorporated in the $\mu_s$ parameter.  In keeping with the known circuitry \cite{veit23}, we assume that the S population receives excitation and inhibits both the E and I populations but receives no inhibition from either inhibitory population. The VIP population is not explicitly modeled here, but rather, will be considered as tonic inhibitory drive to the S population (here, the parameter, $\mu_s$).  In the mean-field model, constant steady states (equilibria) for the system represent {\em asynchronous} activitiy in the spiking network while oscillations (limit cycles) represent synchronous rhythmic behavior for the spiking model. (See Fig. \ref{fig:1}B.)

\begin{figure}
\includegraphics[width=.9\textwidth]{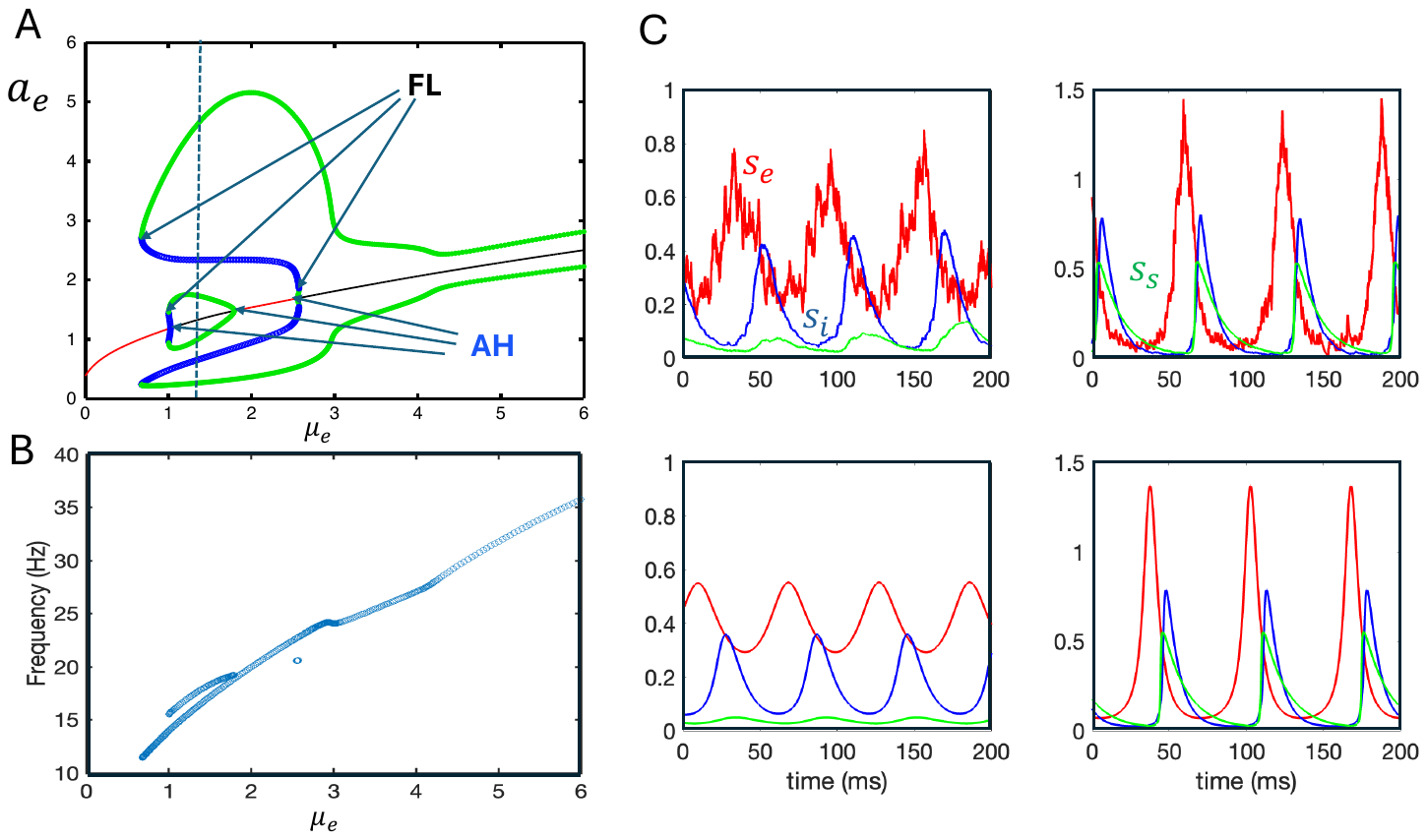}
\caption{Behavior of the mean field model compared to the spiking model. (A) Bifurcation diagram showing the maximum and minimum of $a_e(t)$ as a function of the excitatory drive, $\mu_e$. Solid red (black) lines represent stable (unstable) equilibrium points and thick green (blue) lines are stable (unstable) periodic solutions. Thin vertical line indicates $\mu_e=1.25$ where there are two stable distinct periodic solutions. Six special points are indicated with the thin arrows: Fold of limit cycles (FL) where a stable (green) and unstable (blue) oscillation collide; Andronov-Hopf bifurcation (AH) where the equilibrium changes stability and gives birth to an oscillation. Lightly shaded region depicts parameters where there are two distinct rhythms.(B) Frequency of the oscillations as a function of $\mu_e$ (C) Comparison of the synaptic variables, $s_{e,i,s}$ for the mean-field and the spiking models. Excitatory is red, inhibitory PV is blue, and inhibitory SOM is green.}
\label{fig:2}
\end{figure}

We have chosen our baseline parameter set (Table \ref{tab:par}) such both the $(E,I)$ and $(E,S)$ systems produce limit cycle oscillations when the drive to the excitatory cells, $\mu_e$, is sufficiently strong.  With all parameters fixed ($\lambda=0,\gamma=0.85,\mu_s=-2$), we vary $\mu_e$, the drive to the excitatory neurons and examine what happens in the mean-field model.  Fig. \ref{fig:2}A shows the maximum and minimum values of $a_e(t)$ as $\mu_e$ varies between 0 and 6. There are four classes of solutions: (i) red curves are {\em stable equilibria}; (ii) black curves are {\em unstable equilibria}; (iii) green curves show the maximum and minimum values of {\em stable oscillations}; and (iv) blue curves show maximum and minimum values of {\em unstable oscillations}.  A low excitatory drive, $\mu_e$, only stable equilibria exist corresponding to stable asynchronous activity in the spiking network. Qualitative changes in behavior occur at {\em bifurcations} where, equilibria and limit cycles change stability or appear and disappear. Through almost all of the analysis in this paper, the main two bifurcations are: the {\em Andronov-Hopf} (AH), where an equilibrium changes stability while giving birth to a new limit cycle oscillation ; and the {\em fold of limit cycles} (FL) where stable and unstable limit cycles merge and annihilate.  In panel A, we depict three AH points and three FL points.  The thin vertical line indicates $\mu_e=1.25$, the value used in Fig. \ref{fig:1}. We want to emphasize that each blue or green curve has a maximum and minimum shown in the diagram. 
From this diagram, we can immediately see the range of values of $\mu_e$ where there exist two distinct stable limit cycle oscillations for the same value of $\mu_e$. Since each limit cycle shows the maximum and minimum, any points crossing 4 green lines have two distinct stable oscillations (bi-rhythmicity). The lightly shaded region in panel A shows the primary parameter set where bi-rhythmicity occurs. It is delineated by the region of stability of the smaller limit-cycle which is completely within the region of the large limit cycle.  There is a very small additional region of bi-rhythmicity between the right-most AH and FL.  The larger limit cycle shows a sharp drop in amplitude (coherence in the spiking model) as $\mu_e$ increases from 2 to 3 and a second drop at around $\mu_e=4$.   Fig. \ref{fig:2}B shows the frequency of the rhythms as a function of $\mu_e$ in the same range. The bi-rhythmic region is seen by noting two different frequencies: the smaller amplitude limit cycle has a slightly higher frequency.  The frequency of both rhythms mostly increases with $\mu_e$ and at high drives begins to enter the gamma range (about 35 Hz).  Fig. \ref{fig:2}C compares the mean field values of the synaptic variables, $s_{e,i,s}$ with their counterparts in the spiking model; the left is the larger limit cycle and the right is the smaller (compare Fig. \ref{fig:1}B, bottom). Despite the low number of neurons in the spiking model, the agreement is quite good.  In the smaller limit cycle on the right, the SOM activity (green curve) is very small compared to the PV (blue) and E (green) activity suggesting that the small limit cycle oscillations in panel A are mainly from the $(E,I)$ network. We will see that this is the case shortly.  Because the E cells are firing relatively weakly and $\mu_s=-2$, the S population is only weakly engaged. On the other hand, for the large limit cycle, the S cells are fully engaged and since they also inhibit the I cells, this allows for the E activity to get large despite the strong S to E inhibition.

\begin{figure}
\includegraphics[width=.9\textwidth]{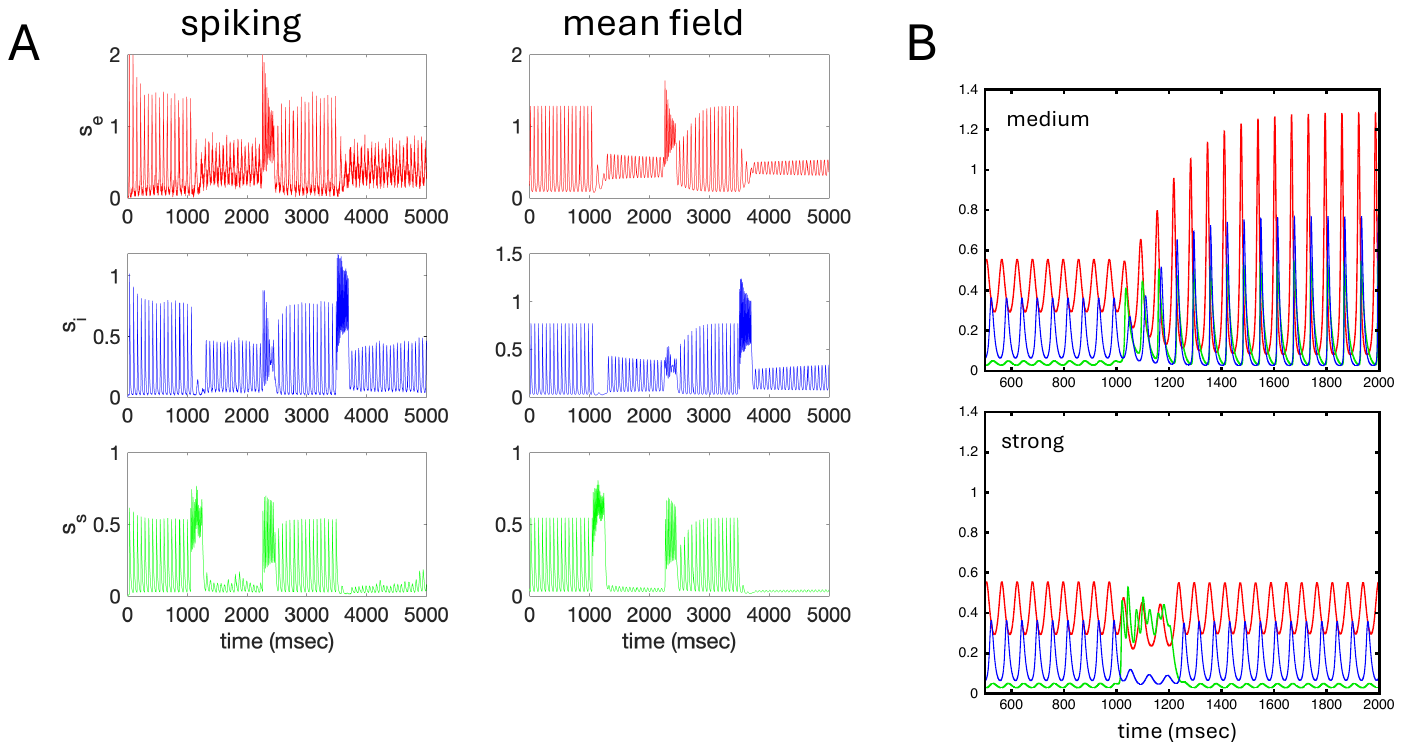}
\caption{(A) Stimuli to different populations of neurons allows for switching between the different oscillations. At $t=1000$ msec, the SOM cells (green) are given a 200 msec input which switches from the large oscillation to the smaller one; at $t=2050$ msec, the E cells (red) are given a stimulus for 200 msec to switch back to the large oscillation; at $t=3500$ the PV cells (blue) are given a 200 msec stimulus switching from the large to the small oscillation. (B) Stimuli of appropriate amplitude to the S population can switch from the small oscillation to the large (top), but if the stimulus is too large, no switch occurs.}
\label{fig:3}
\end{figure}

Since our network is bi-rhythmic, we should be able to switch between states by stimulating one or more of the populations. In Fig. \ref{fig:3}A, we show how 200 msec input pulses can switch between states both in the mean field and in the spiking network.  We initialize both networks so that they both lie on the large limit cycle.  At $t=1050$ we provide a strong pulse of activity to the S cells, (green burst). This transiently suppresses both the E and I cells, with suppression of the E cells sufficent to push them down to the smaller limit cycle.  At $t=2000$, we stimulate the E population which switches it back to the large limit cycle. Finally, at $t=3500$, we stimulate the I population which suppresses the E cells enough to bring the system to the small limit cycle.   Interesting, because the S population inhibits both I and E, it is possible to switch the network from the small to large oscillations by appropriately stimulating the S cells. In Fig. \ref{fig:3}B, we apply a 200 msec stimulus with amplitude 1 to the S population. This causes the S activity (green) to increase and at the same time suppresses the I activity (blue), but the E activity (red) is not suppressed that much. The suppression of I allows the E to grow enough to reach the large oscillation (top panel).  However, if the input to S is too strong (bottom panel) then, while the I population is strongly suppressed, so too is the E population and the switch to the large oscillation does not occur.

So far, we have shown that having two distinct populations of inhibitory interneurons provides a simple mechanism for the appearance of multiple beta frequencies in a coupled network. Some natural questions to ask are: how necessary are two populations; how robust are these behaviors; and are there other types of dynamics.  To answer these questions, we have introduced two parameters in the model equations: $\gamma$ and $\lambda$, which scale the inhibition from the I and S cells respectively (see Eq. (\ref{eq:eisoa}). For example, increasing $\gamma$ has the effect of proportionally increasing the I to E and I to I inhibition. Similarly, setting, $\lambda=0$ removes the S inhibition from the circuit.  In Fig. \ref{fig:2}A, we saw that the stable states of the system are determined by the AH and FL bifurcations as these are where oscillations are born and where they die.  Thus, in the next few figures, we will co-vary the excitatory drive, $\mu_e$ and the scaling factors, $\lambda$ and $\gamma$.  Specifically, for each value of, say, $\lambda$, we will track the values of $\mu_e$ where there are AH and FL bifurcations.  This will parcel the $(\mu_e,\lambda)$ plane into regions with different qualitative behaviors. The ensuing diagrams are called {\em two-parameter bifurcation diagrams} and we will also replot them in a more intuitive manner where the different types of stable behavior are summarized.

\begin{figure}
\includegraphics[width=.9\textwidth]{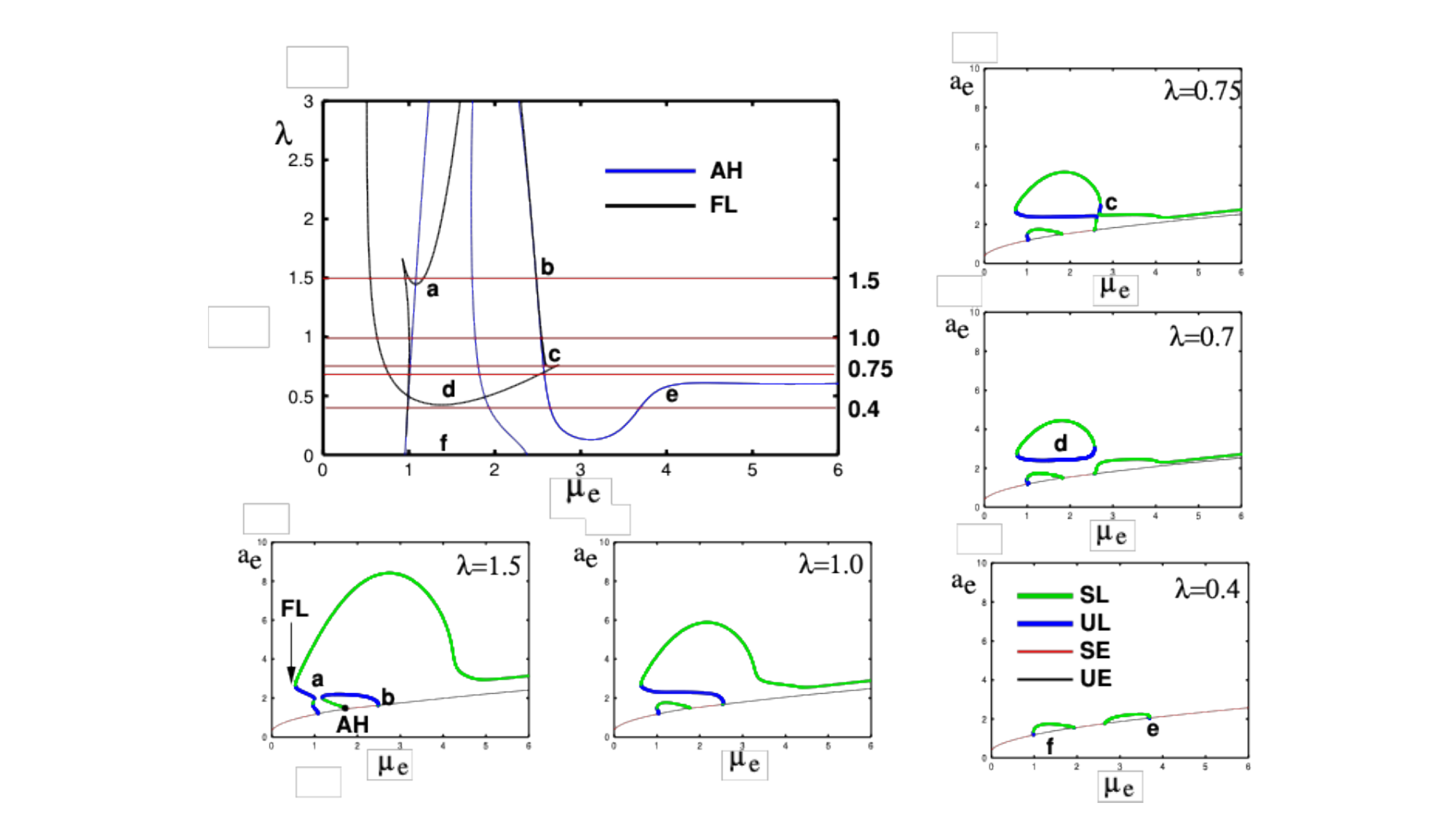}
\caption{Dependency of dynamics and multi-rhythmicity on SOM ($\lambda$). The number of oscillations is organized around the appearance and disappearance through collisions of stable and unstable oscillations (FL, shown as black curves) and the emergence and disappearance of oscillations from equilibria (AH, shown as blue curves). As the degree of SOM inhibition ($\lambda$) changes the dependence of the mean-field system on the excitatory drive ($\mu_e$) changes qualitatively. Crossing the black and blue curves changes the number of stable equilibria and oscillations. The letters a-f in the $(\mu_e,\lambda)$ diagram are shown at their corresponding values in the $(\mu_e,a_e)$ diagrams surrounding the main diagram. SL (UL):stable (unstable)oscillation; SE (UE): stable (unstable) equilibrium.  The label {\bf d} shows an example of an {\em isola of oscillations} where solutions are not connected to the main branch of equilibria. Once $\lambda$ falls below about 0.45, there is no bi-rhythmicity. }
\label{fig:4}
\end{figure}

Fig.~\ref{fig:4} shows the behavior in the $(\mu_e,\lambda)$-plane to see the consequences of varying the degree of S inhibition. To better clarify this diagram, we also depict diagrams of $a_e$ vs $\mu_e$ for different values of $\lambda$. Starting with $\lambda=0$ (just I and E cells), as $\mu_e$ increases, the first blue (AH) line (at $\mu_e=0.98$) is crossed  giving birth to a small amplitude limit cycle which persists until the second blue curve is crossed and this limit cycle is lost. (We do not show this plot, but it is very similar to the $\lambda=0.4$ case at the label {\bf f}.) As we increase $\lambda$, it next crosses a pair of AH (blue) curves producing another small amplitude stable oscillation.  This is shown in the $\lambda=0.4$ panel. The point {\bf e} indicates the right-most AH point. A very interesting bifurcation occurs as we further increase $\lambda$ past around 0.45 near the point {\bf d}. The FL (black) curves are crossed twice leading to an {\em isola} of limit cycles. That is, there is an isolated pair of stable and unstable limit cycles which can be seen in the $\lambda=0.7$ panel. The small amplitude oscillation for $\mu_e$ roughly between 1 and 2 persists, but the other small amplitude oscillation ({\bf e} in the $\lambda=0.4$ panel) now extends past $\mu_e=6$. This can be read off the main diagram since the line $\lambda=0.7$ is above the AH curve {\bf e}.  Note that there is now bi-rhythmicity between the small limit cycle and the isola. As $\lambda$ further increases, the isola merges with the right-most small amplitude oscillation ({\bf c} in the $\lambda=0.75$ panel). Further increases in $\lambda$ lead to dynamics that are qualitatively like those in Fig. \ref{fig:2}A and the $\lambda=1$ panel in the present figure. Increasing $\lambda$ to 1.5 merges the big limit cycle and the small limit cycle on the left and the points {\bf a,b}.

\begin{figure}
\includegraphics[width=.9\textwidth]{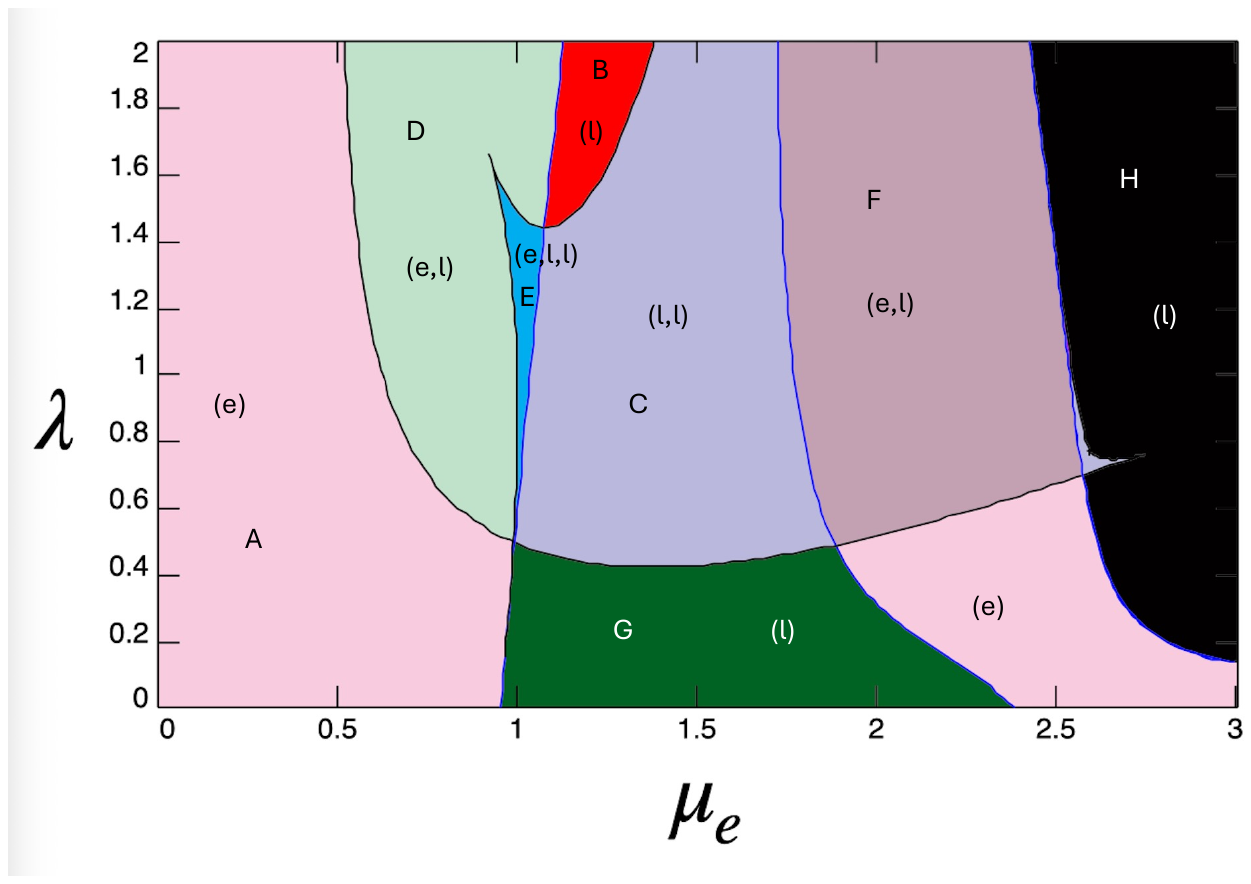}
\caption{Two-parameter phase-diagram showing how the curves in Fig. \ref{fig:4} divide the $(\mu_e,\lambda)-$ plane into regions with specific numbers of attractors. The letters in parentheses indicate the distinct stable dynamics. For example, in region E, (e,l,l) means there is a stable equilibrium and two stable limit cycles. Regions B,G, and H all contain just one stable limit cycle. Unstable dynamics are not indicated in the figure. As in Fig. \ref{fig:4}, the regions are separated by black (FL) and blue (AH) curves.  }
\label{fig:5}
\end{figure}

To better clarify the qualitative states in the network, we use the AH and FL curves to divide the $\mu_e-\lambda$ plane into distinct regions where there are different stable states. Fig. \ref{fig:5} shows this diagram using different colors to aid in separating the regions. Within the labeled regions, we indicate the possible {\em stable} states in parentheses.  For example, in region E, we have (e,l,l). This means that there are two stable limit cycles (l,l) and one stable equilibrium (e).  In regions B,G,H, there is just one stable limit cycle, but we have colored them differently as they are separate ``branches'' of oscillations. In region B the sole stable oscillation is a large amplitude one and corresponds to the large limit cycles seen in the $\lambda=1,1.5$ panels in Fig. \ref{fig:4}. In region G the sole oscillation corresponds to the small amplitude rhythm shown at the point {\bf f} where $\lambda=0.4$ in Fig. \ref{fig:4}, while region H is the small amplitude oscillation that is seen in all the panels in Fig. \ref{fig:4} when $\mu_e$ is large.  The large region C shows values of $(\mu_e,\lambda)$ where there is birhythmicity.  In region A, only a single stable equilibrium occurs. We can use the AH,FL curves to understand transitions from one region to another. For example, in the transition from C to F [(l,l) to (e,l)], the blue AH curve is crossed and a small-amplitude oscillation is absorbed by an equilibrium which becomes stable.  In the transition from C to B [(l,l) to (l)], a FL curve is crossed indicating that a stable and unstable limit cycle have collided and disappeared.  Finally, in the transition from C to G, the isola is lost, leading to just a small amplitude oscillation.

\begin{figure}
\includegraphics[width=.9\textwidth]{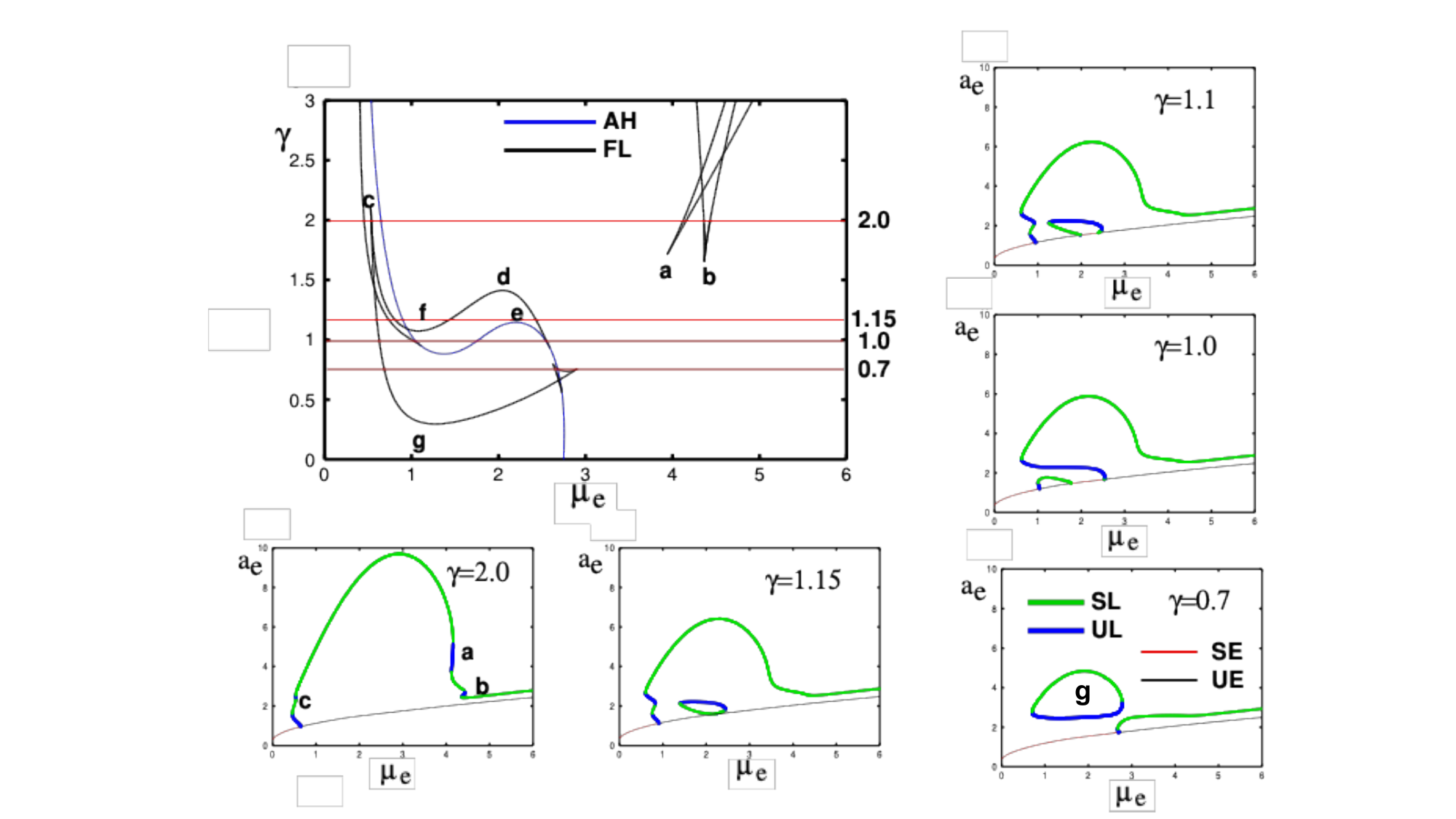}
\caption{Dependency of dynamics and multi-rhythmicity on PV ($\gamma$). See Fig. \ref{fig:4} for explanation of curves. If $\gamma$ falls below about $0.3$, multi-rhythmicity is lost. }
\label{fig:6}
\end{figure}

\begin{figure}
\includegraphics[width=.9\textwidth]{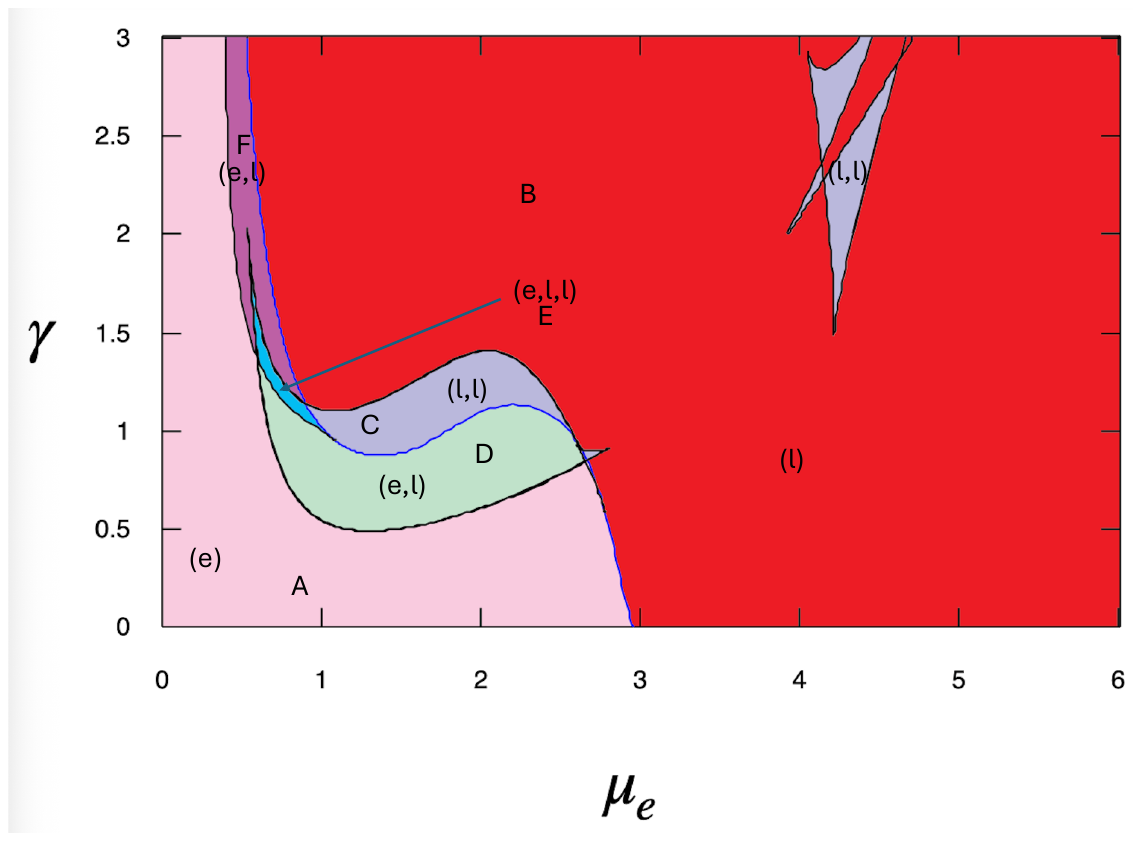}
\caption{Two-parameter phase-diagram showing how the curves in Fig. \ref{fig:5} divide the $(\mu_e,\gamma)-$ plane into regions with specific numbers of attractors. See Fig. \ref{fig:5} for explanation. }
\label{fig:7}
\end{figure}

Fig.~\ref{fig:6} illustrates the behavior in the $(\mu_e,\gamma)$-plane, focusing on the consequences of varying the degree of I-inhibition. To further clarify this diagram, we also present plots of $a_e$ versus $\mu_e$ for different values of $\gamma$. Starting with $\gamma = 0$ (where only S and E cells are present), as $\mu_e$ increases, the system crosses the first blue AH (Andronov-Hopf) bifurcation line at $\mu_e = 2.7$, giving rise to a small amplitude limit cycle. Although this plot is not shown, it closely resembles the $\gamma = 0.7$ case without the isola. As $\gamma$ increases, an isola emerges when a horizontal branch crosses two fold limit cycle (FL) curves around $\gamma = 0.4$ in the two-parameter diagram. As $\gamma$ increases, a pair of AH bifurcations occur, leading to a small branch of limit cycles (seen in the $\gamma=1$ diagram for $\mu_e$ roughly between 1 and 2). The isola at $\gamma=0.7$ merges with the branch of oscillations ($\mu_e>2.75$) to form one large branch (labeled {\bf g} in the $\gamma=1$ diagram).   Between $\gamma=1$ and $\gamma=1.1$, the large limit cycle collides with the small branch (point {\bf f}) to form one large branch with a smaller branch coming from the pair of AH bifurcations (point {\bf e}). As $\gamma$ increases slightly beyond $\gamma = 1.1$ the AH blue curve is crossed (point {\bf e} in the $\mu_e-\gamma$ diagram to produce another isola (point {\bf e} in the $\gamma=1.15$ diagram).   With continued increases in $\gamma$, the isola gradually shrinks and eventually disappears near $\gamma = 1.5$. Birhythmicity is observed for $\mu_e$ in the interval $[1, 2]$ when $\gamma$ ranges approximately from $0.8$ to $1.5$.

In  Fig.~\ref{fig:7} we again partition the $(\mu_e, \gamma)$-plane into distinct regions. Region E signifies two stable limit cycles and one stable equilibrium. Regions B, D, and F each contain a single stable limit cycle, but they are distinguished by color since they represent different oscillatory branches, not all of which coexist with equilibrium points. In Region B, the stable oscillation is a large amplitude limit cycle that subsequently decreases in size; this is consistently seen across all $\gamma$ panels in Fig.~\ref{fig:6}. Region D features a large amplitude rhythm coexisting with a stable equilibrium, typically occurring for $\gamma$ values from 0.5 to 1.3. In contrast, Region F contains a stable limit cycle whose amplitude increases with $\mu_e$ and is visible in the panels where $\gamma$ ranges from approximately 1.3 to 2 in Fig.~\ref{fig:6}. Regions C and E both exhibit birhythmicity; however, Region E also includes a stable equilibrium. Region A is characterized by the presence of a single stable equilibrium.  The AH and FL curves can be used to understand the transitions between these regions. For example, the transition from Region C to Region B (from (l, l) to (l)) occurs when the system crosses a black FL curve, leading to the disappearance of a small amplitude oscillation, which is absorbed by an unstable limit cycle. Similarly, for the transition from Region C to Region D occurs when a small amplitude stable limit cycle collides with an equilibrium at the AH bifurcation (blue curve).

\begin{figure}
\includegraphics[width=.9\textwidth]{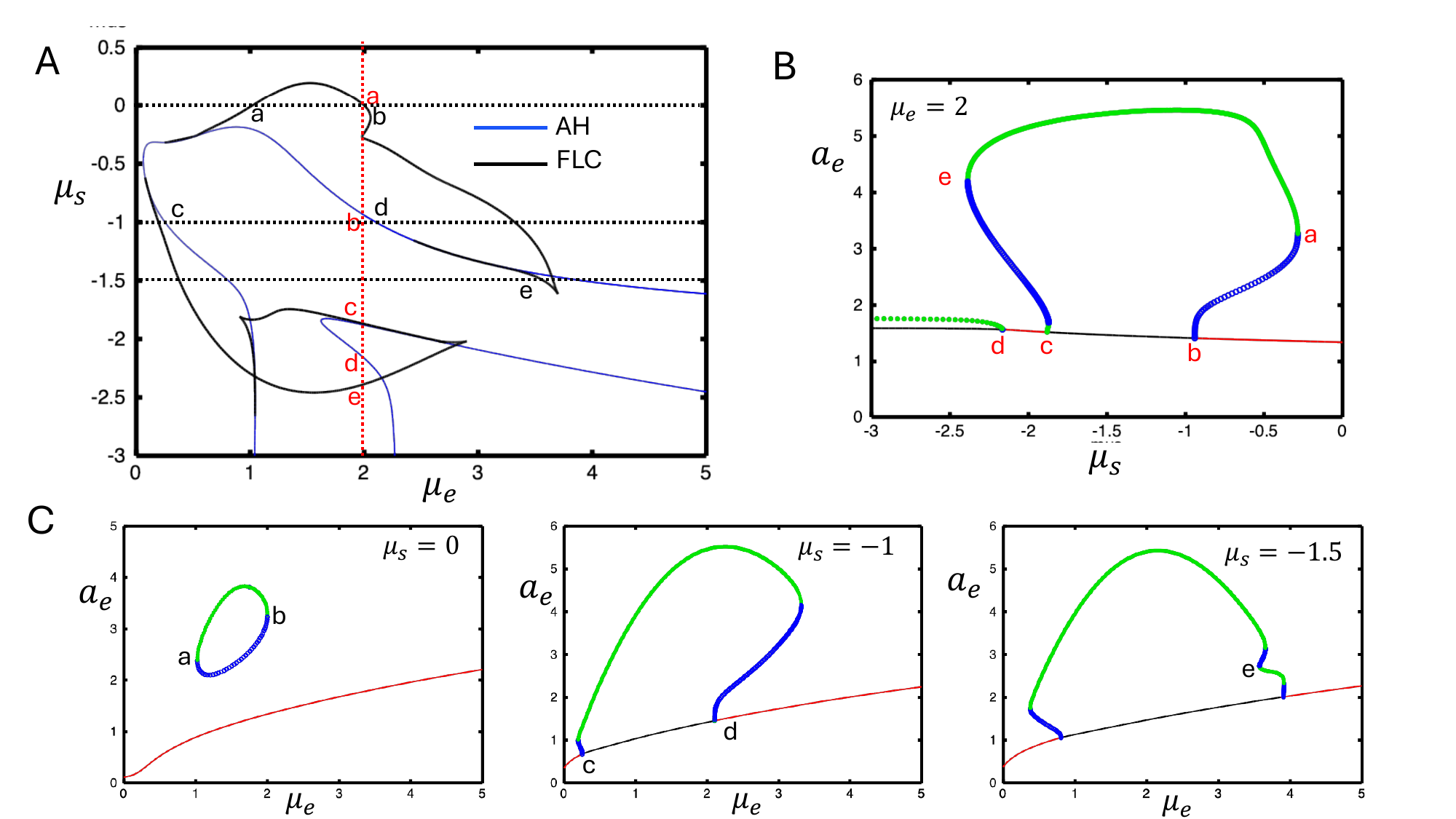}
\caption{Modulation of SOM by VIP ($\mu_s$) has important effects on the dynamics.  A. As in Figs. \ref{fig:4},\ref{fig:5} we show the collisions of stable and unstable limit cycles (FL) in black and the emergence of oscillations from equilibria (AH) in blue. B. With $\mu_e=2$ fixed (red dashed line in A), we study the behavior as $\mu_s$ is increased from -3. Red letters correspond  to points in A along the vertical dashed red line. C. Dependence on $\mu_e$ for different values of $\mu_s$.    }
\label{fig:8}
\end{figure}

\begin{figure}
\includegraphics[width=.9\textwidth]{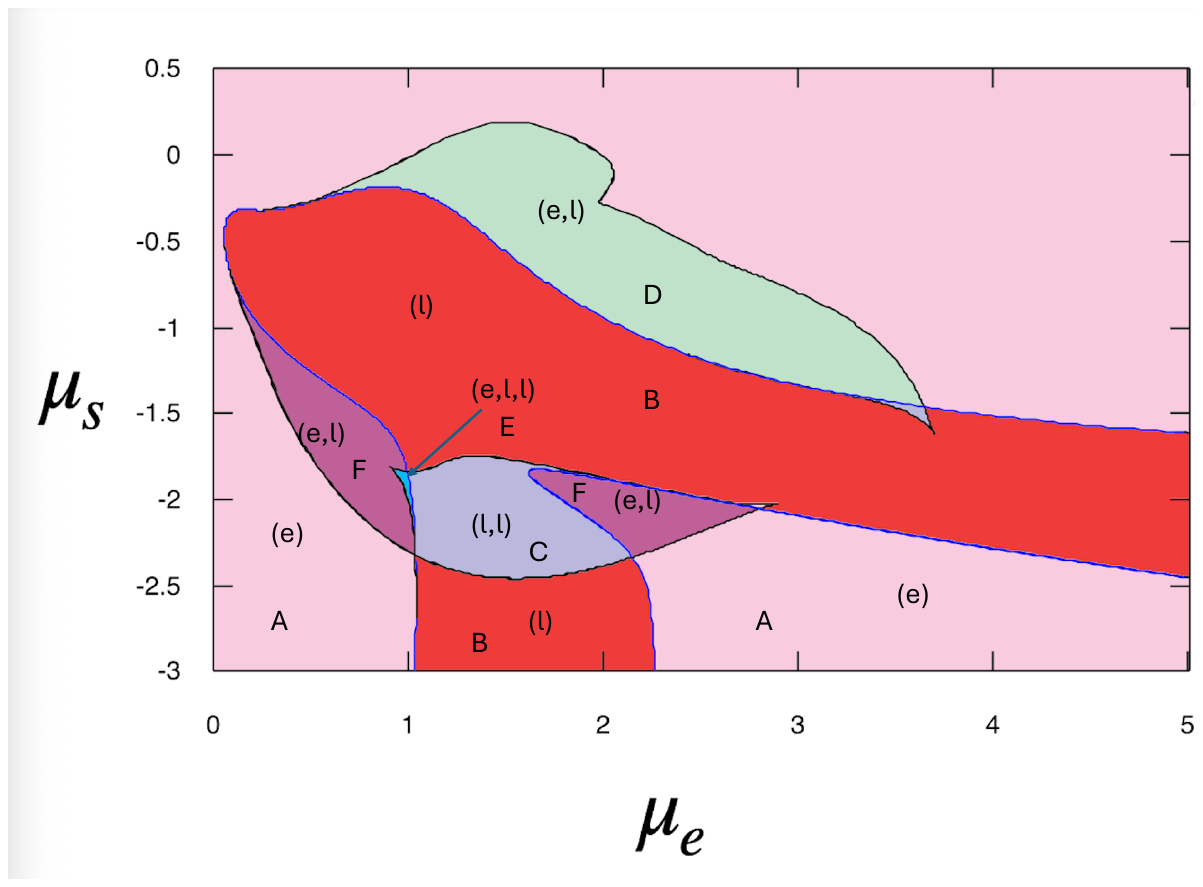}
\caption{Two-parameter phase-diagram showing how the curves in Fig. \ref{fig:8} divide the $(\mu_e,\mu_s)-$ plane into regions with specific numbers of attractors. See Fig. \ref{fig:5} for explanation.}
\label{fig:8b}
\end{figure}

The primary role of VIP neuronal input is to modulate the SOM neurons \cite{veit23}, thus, we next explore how altering the input in the SOM neurons ($\mu_s$) affects the presence and multiplicity of rhythms in the EIS network. We combine the inhibition from the VIP neurons with the tonic drive to the SOM cells into the parameter $\mu_s$, so that more negative values of $\mu_s$ correspond to stronger VIP input. 
In Figs. \ref{fig:8},\ref{fig:8b} we show the $(\mu_e,\mu_s)$ two-parameter diagram with sample one-parameter diagrams at different levels of $\mu_s$ and also the division of parameters into different regimes of qualitatively similar behavior.
As with the case where we varied the strength of the SOM inhibition onto the E and I populations (Figs. \ref{fig:4},\ref{fig:5}), varying $\mu_s$ has very similar effects.  By fixing $\mu_e$, we can follow the effects of VIP inhibition, as shown, for example in Fig. \ref{fig:8}B where $\mu_e=2$.   At high VIP inhibition ($\mu_s$ very negative), there are only small amplitude oscillations driven by the E/I subnetwork (below the point labeled d). Decreasing VIP inhibition produces an isola (region C in Fig. \ref{fig:8b}) and bi-rhythmicity as the SOM population can now interact with the E population to generate rhythms (between the points labeled c and d in Fig. \ref{fig:8}B). Further increases in $\mu_s$ (decreasing VIP inhibition) increases the activity of the SOM neurons and kills the small amplitude oscillation (point d). At $\mu_s=-1$ (point b), the asynchronous state is stabilized leading to bistability between the large amplitude rhythm and the asynchronous state.  Finally when VIP inhibition is near 0 ($\mu_s=-0.3$, point a), the SOM is so active that all that remains is the stable asynchronous state.

\begin{figure}
  \includegraphics[width=0.9\textwidth]{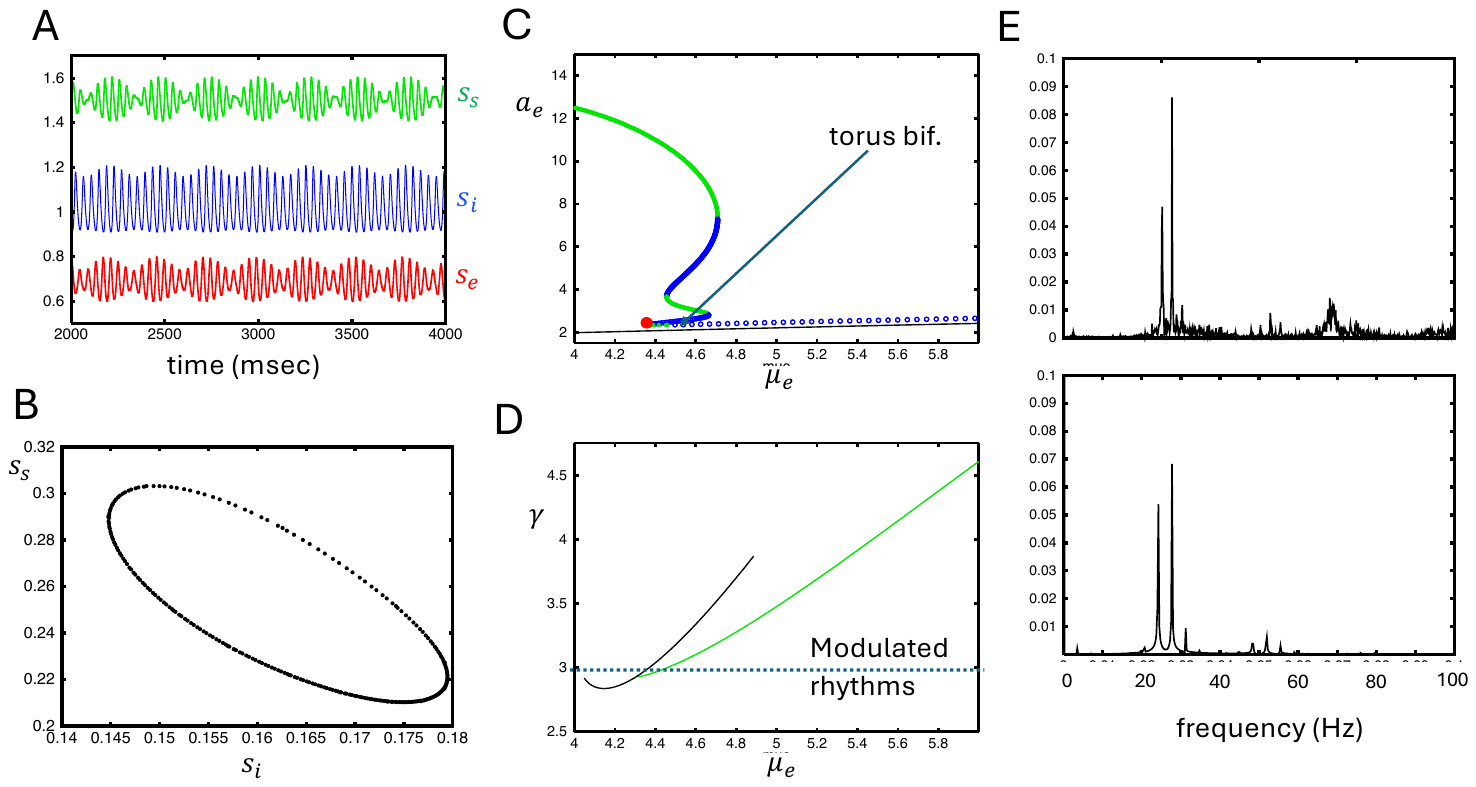}
  \caption{Quasi-periodic dynamics. (A) Synaptic dynamics (vertically translated) for $\gamma=3,\mu_e=4.8$ showing amplitude modulation; (B) values of $(s_i,s_s)$ whenever $s_e(t)$ crosses 0.7 from below; (C) one-parameter diagram showing the appearance of the {\em torus bifurcation} (onset of modulated rhythms); solid red circle shows a fold of limit cycles producing oscillations that eventually become the quasi-periodic solutions shown in (A,B); (D) $(\mu_e,\gamma)-$diagram showing (green line) where modulated solutions arise. They terminate at the left when the green line runs into the fold (black curve); (E) Power spectrum of $s_e(t)$ for the spiking model (top) and the mean field (bottom) dynamics showing peaks at about 24 and 27 Hz.}
  \label{fig:tor}
\end{figure}

We have shown that over a wide range of synaptic strengths of the SOM and PV cells, our network is capable of producing two {\em distinct} rhythms that are close in frequency but distinct in the degree in which the SOM cells participate and in the degree of coherence (c.f. Fig. \ref{fig:2}C).  One common feature in local field potential (LFP) recordings are multiple peaks in the power spectrum. A simple mechanism for these multiple peaks is that there are just different independent populations and the LFP represents their summed activity.  A more interesting mechanism is that the multiple frequency peaks are all part of the same circuit and represent {\em quasiperiodic} behavior. Mathematically, quasi-periodic behavior can arise when a limit cycle loses stability to a {\em torus bifurcation}.  We have found that setting $\gamma>3$ (very strong PV inhibition) can result in this kind of behavior for our network.  In Fig. \ref{fig:tor}A, we show the synaptic activity of the E,I, and S populations for $\mu_e=4.8,\gamma=3$. There is a high frequency rhythm whose amplitude is slowly modulated with a frequency of 3.75 Hz indicative of quasiperiodic behavior.  To further check whether this is the case, in panel B, we plot a dot in the $(s_i,s_s)-$plane each time $s_e$ crosses the value 0.7 from below.  The result of this plot is an almost perfect ellipse indicating that dynamics is filling up a two-dimensional torus.  In  panel C, we draw a bifurcation diagram with $\gamma=3$ as $\mu_e$ increases. We only show a small window of the diagram as we want to focus the reader's attention in the region where $\mu_e\approx 4$. As $\mu_e$ increases past 4, there is a FL at $\mu_e\approx4.66$; then as $\mu_3$ decreases another FL occurs at $\mu_e\approx 4.45$. As $\mu_e$ again increases a third FL appears at $\mu_e\approx4.71$ and finally as $\mu_e$ decreases, a fourth FL appears (marked by the red circle) at $\mu_e\approx4.40$. This produces a very small amplitude stable limit cycle that loses stability at $\mu_e\approx4.46$ to a {\em torus bifurcation} marked by the arrow.  For $\mu_e$ larger than this value, we find stable quasi-periodic behavior. With $\gamma$  as a second parameter, we show, in panel D, the range in $(\mu_e,\gamma)$ where there exists quasiperiodic or modulated rhythms; to the right and below the green line marked TR. This curve terminates on the left when it hits the FL curve (black) that corresponds to the FL with the red dot in panel C.  In panel E, we show the power spectrum for $s_e(t)$ for $\mu_e=4.8,\gamma=3$ for both the mean-field model (bottom) and the full spiking model (top).  There are two major peaks at 24.5 and 28.25 Hz and a smaller peak at 3.75 Hz that occurs at the difference of these two frequencies and is resposible for the slow modulation (or beating).  In the spiking model there is a rough peak at about 65 Hz which is due to the spiking of individual neurons in the finite population and as $N\to\infty$, this will disappear.

\begin{figure}
\includegraphics[width=.9\textwidth]{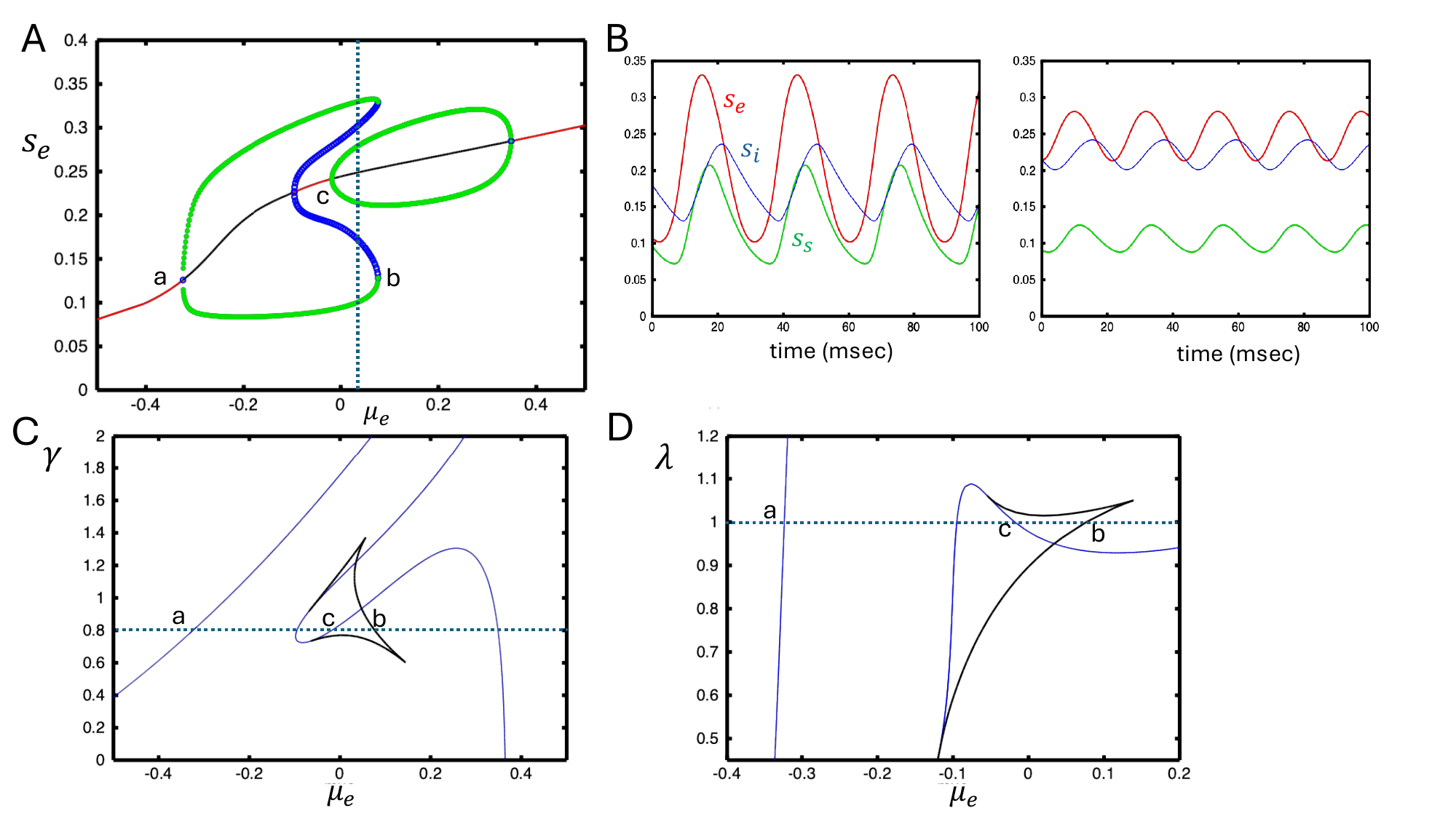}
\caption{The three-dimensional firing rate approximation of the 9D meanfield model also shows bi-rhythmicity for narrow ranges of parameters. (A) Bifurcation dfiagram as $\mu_e$ varies. Points (a,c) show the onset of an oscillation through an AH. Point (b) shows the collision of a stable and unstable oscillation (FL). For $c<\mu_e<b$ there is birhythmicity. (B) The two limit cycle oscillations for $\mu_e=0.05$. (C,D) two-parameter diagrams showing the FL (black) and AH (blue) with $\mu_e$ varying and either $\lambda$ or $\gamma$.  Dashed line corresponds to the values in A. Bi-rhythmicity occurs within the cusped regions in these diagrams (see arrows). }
\label{fig:9}
\end{figure}

We close this paper with an analysis of a simpler model, similar to the Wilson-Cowan firing rate model.  If we let the membrane time constants $\tau_{m,k}$ ($k\in\{e,i,s\}$) go to zero, then the equations for $a_k,b_k$ reduce to algebraic equations for which we can solve:
\begin{equation}
  \label{eq:frdef}
a_k = \left[\frac{1}{2}\left(I_k+\sqrt{I_k^2+\Delta_k^2}\right)\right]^\frac{1}{2}
\end{equation}
where $I_k=\mu_k+g_{ek}s_e+g_{ik}s_i+g_{sk}s_s$ is the total input into each population.  We then obtain a set of three equations for $s_e,s_i,s_s$:
\begin{eqnarray}
  \tau_{e} \dot{s_e} &=& -s_e+a_e/\pi \nonumber \\
  \label{eq:fr}
  \tau_{i} \dot{s_i} &=& -s_i+a_i/\pi \\
  \tau_{s} \dot{s_s} &=& -s_s+a_s/\pi, \nonumber
\end{eqnarray}
where the $a_k$ are defined in Eq. (\ref{eq:frdef}).   By looking separately at the $(s_e,s_i)$ and $(s_e,s_s)$ systems we were able to find a set of parameters where there was a range of drives to the E population that lead to birhythmicity.  Fig. \ref{fig:9}A shows the bifurcation diagram as $\mu_e$ increases.  At the point labeled {\bf a} an AH bifurcation occurs and a stable oscillation emerges from this that terminates at a FL at point {\bf b}. At a value of $\mu_e$ lower that this point, another limit cycle emerges from an AH bifurcation (point {\bf c}) and persists until $\mu_e$ becomes sufficiently large.  Birhythmicity exists for $\mu_e$ between {\bf c} and {\bf b}.  Panel B shows $s_e(t),s_i(t),s_s(t)$ for $\mu_e=0.04$  for both the ``big'' and the ``small'' oscillations. The left oscillation has a frequency of 34 Hz and the  
right, a frequency of 46 Hz, both higher than the oscillations seen in the full mean-field model.  Fig. \ref{fig:9}C,D show the corresponding two-parameter diagrams showing the FL (black) and AH (blue) bifurcation curves. In the small regions inficated with the arrows, there is birhythmicity.  While the bifurcation structure is considerably simplified compared to the full mean-field, there are qualitative similarities. The limit cycle bifurcating at the lowest value of $\mu_e$ is primarily due to the PV cells as $\mu_s$ is very negative. At higher values of $\mu_e$, the SOM cells are engaged and mainly responsible for the oscillations.

\section*{Discussion}

In this paper, we have shown that adding multiple types of inhibitory neurons greatly extends the types of oscillatory dynamics that are possible in cortex-like circuits. Starting with a spiking network composed of coupled quadratic integrate-and-fire neurons, we found that for certain ranges of parameters, there was multi-rhythmicity. That is there were two distinct rhythmic states and we can switch between them with appropriate stimuli to the excitatory or inhibitory classes. The range of parameters over which there is bi-rhythmicity is fairly broad, so it may be possible to find this kind of dynamics in {\em in vitro} preparations. Since it is possible to selectively stimulate different populations of neurons with optogenetics, it might even be possible to experimentally switch between rhythms.

To facilitate the analysis of this oscillatory circuit, we employed an exact mean-field reduction which produced a nine-dimensional set of ordinary differential equations amenable to a thorough bifurcation analysis. We examined the contribution of the two different classes of inhibition by fixing the conductance parameters and then varying the excitatory drive while at the same time scaling the strength of each inhibitory population onto the other populations (I to E and I to I or S to E and S to I). Figs. \ref{fig:5} and \ref{fig:7} show that multi-rhythmicity requires contributions from both classes of inhibition. The region of multi-rhythmic behavior occurs over a larger range of SOM (S) interactions than with PV (I) as can be seen in the much larger purple area in Fig. \ref{fig:5} than in Fig. \ref{fig:7}.  We also found small areas where there is tri-stability with two stable rhythms and a stable equilibrium. Oscillations appear and disappear via two different mechanisms: (1) Hopf bifurcation where a new oscillation arises from or is absorbed into an equilibrium; (2) Folds of limit cycles when an unstable and stable oscillation pair emerges or is collides.  Related to this, we found regions where there are isolated branches of limit cycles such as in Fig.~\ref{fig:4} ($\lambda=0.7$)  or Fig. \ref{fig:8} ($\mu_s=0.0$).   This mechanism could provide an alternate model for the multiple beta rhythms studied in \cite{rkc}. 

It has been shown that VIP inputs into SOM cells have an important modulatory effect on the circuit \cite{bos20,edwards24}, thus we also looked at the effects of VIP inputs onto the SOM. Here, we treated the input as a constant negative drive and as would be expected, at high VIP input ($\mu_s$ very negative), we essentially remove S from the circuit and multirhythmicity disappears.  If the VIP input is too weak (that is, $\mu_s$ increases), then birhythmicity disappears.

With strongly enhanced PV coupling strength (roughly 3-fold the default), we found a regime where theres is {\em quasi-periodicity}, Fig. \ref{fig:tor} manifested as a low-frequency (3 Hz) amplitude modulation of higher frequency oscillations (25-27 Hz) and two close peaks in the power spectrum.  This type of behavior has not been seen in simple E-I networks to our knowledge and could provide a mechanism for theta-gamma coupling \cite{lisman2013}.  These appear with high drive to the E cells and strong PV  inhibition. Interestingly, there is a small region of drives to the E cell where there is quasi-periodicity and {\em two additional stable oscillations} (Fig. \ref{fig:tor} C).

By letting the membrane time constants of all three cell types go to zero, we were also able to produce a three variable ``Wilson-Cowan'' type model which has a narrow rnage of parameters where there are two stable oscillations (Fig. \ref{fig:9}).

The present study is similar to a recently published paper \cite{FMM}, where the authors employed a network of exponential integrate-and-fire neurons with E,S, and I cells.  One major difference is that their I cells have the same synaptic decay time, while in our model the SOM (S) cells decay slower than the PV (I) \cite{edwards24}. Like \cite{FMM}, the frequency of our rhythms was in the lower gamma range and closer to beta (20-30 Hz), however, with enough drive, we are able to get higher frequency oscillations such as seen in E/I networks \cite{borgers05}. The ``noise'' in \cite{FMM} arises from Poisson inputs into the cell types. In our study, the source of noise comes from the heterogeneity in the external drives which was used so that we could create a mean-field model. Recent work \cite{goldobin21} suggests some generalizations of the mean-field equations we studied here are possible when there is Gaussian noise. Like \cite{FMM}, the spike-phase relationships between the SOM and PV cells with the LFP are tighter than those observed experimentally (cf Fig.1 in \cite{FMM}), however, in both their study and ours, PV cells fire closer to the peak of the LFP than do SOM cells; this is particularly apparent in the small limit cycle (see Fig. \ref{fig:1b}).  Because they use a high-dimensional spiking model that has no simple mean-field description, they explored a limited set of parameter values. However, they also interchangeds the SOM and PV connectivity to better tease out the contributions of the two populations. The recent paper \cite{bos20} used a firing rate formulation and was aimed at the analysis of gain modulation rather than rhythmicity. \cite{edwards24} found regions of oscillatory dynamics at 20-30 Hz when they drove the SOM populations in their spiking model based on the exponential integrate-and-fire model.  Their model included VIP neurons and because they were driven by the E cells, they are an integral part of the circuit.  In our simplified circuit, VIP is just an inhibitory drive to the SOM population.  Unlike our model and \cite{FMM}, \cite{edwards24} employed a spatially distributed model. A major difference betwween our results and theirs is that they fail to see oscillations in when the SOM population is strongly inhibited. Without reduced SOM, we still have regions when there are rhythms such as Fig. \ref{fig:4} ($\lambda=0.4$).

A natural question that arises is whether there are any computational advantages to having a system with multiple stable rhythms. Since it has been hypothesized that oscillatory neural activity  can organize and synchronize local assemblies of neurons \cite{kopell11,bw12,fernandez}, having multiple stable oscillations might allow for handling multiple input streams in parallel.

In the present study, our E cells do not have adaptation. In a recent paper \cite{grishma} we showed that inhibition controlled the sparsity of firing in E/I networks and that adaptation only played a role when the inhibition was greatly reduced. In the present study, we have two types of inhibition so that even if one type is reduced, the other will compensate.   Adaptation was incorporated in \cite{FMM} with a time constant of 500 msec. Since the time scales of our oscillations are nearly ten-fold faster, the addition of adaptation would effectively be a negative constant current.  Our neurons were point neurons and SOM inhibition typically targets dendrites rather than soma. We have given our SOM cells a longer synaptic decay constant to approximate the electrotonic effects of dendritic inputs. It could in principle be possible to add a second compartment to the E cells, but it is unclear whether one could then create a mean field model.

\section*{Materials and methods}

 \begin{table}[h!]
    \centering
    \begin{tabular}{|l|l||l|l||l|l|}
        \hline
        $\mu_e$ & 1.25 & $\mu_i$ & -0.5 & $\mu_s$ & -2 \\
        \hline
        $g_{ee}$ & 1.5 & $g_{ei}$ & 2.0 & $g_{es}$ & 4.25 \\
        $g_{ie}$ & 1.0 & $g_{ii}$ & 0.5 & $g_{is}$ & 0 \\
        $g_{se}$ & 2.0 & $g_{si}$ & 0.5 & $g_{ss}$ & 0 \\
        \hline
        $t_{me}$ & 20 & $t_{mi}$ & 10 & $t_{ms}$ & 10 \\
        $\tau_{e}$ & 2.0 & $\tau_{i}$ & 7.5 & $\tau_{s}$ & 15 \\
        \hline
        $\Delta_{e}$ & 0.1 & $\Delta_{i}$ & 0.1 & $\Delta_{s}$ & 0.1 \\
        \hline
        $p$ & 0.95 & $\lambda$ & 0.85 &  $\gamma$ & 1  \\
        \hline
    \end{tabular}
    \caption{Default parameter values for the network model.}
    \label{tab:par}
\end{table}

The spiking model consists of three populations of 400 quadratic integrate and fire neurons:
\begin{align}
  \label{eq:spike}
v_{ej}'&=\frac{1}{\tau_{me}}\left[i_e(t)+(v_{ej})^2+\mu_e+\Delta_e\zeta_{je}+g_{ee}s_e-g_{ie}s_i-\lambda g_{se}s_s\right]\\
v_{ij}'&=\frac{1}{\tau_{mi}}\left[i_i(t)+(v_{ij})^2+\mu_i+\Delta_i\zeta_{ji}+g_{ei}s_e-g_{ii}s_i-\lambda g_{si}s_s\right]\\
v_{sj}'&=\frac{1}{\tau_{ms}}\left[i_s(t)+(v_{sj})^2+\mu_s+\Delta_s\zeta_{js}+g_{es}s_e-g_{is}s_i-\lambda g_{ss}s_s\right]\\
s_e'&=-\frac{s_e}{\tau_e}+-\frac{1}{400}\sum_{j=1}^{400}\left(\frac{\tau_{me}}{\tau_e}\delta(t-t_{ej}\right)\\
s_i'&=-\frac{s_i}{\tau_i}+-\frac{1}{400}\sum_{j=1}^{400}\left(\frac{\tau_{mi}}{\tau_i}\delta(t-t_{ij}\right)\\
s_s'&=-\frac{s_s}{\tau_s}+-\frac{1}{400}\sum_{j=1}^{400}\left(\frac{\tau_{ms}}{\tau_s}\delta(t-t_{sj}\right).
\end{align}
where $v_{zj}$ is the membrane potential of the $j^{th}$ neuron ($j=1,\ldots,400$) in population $z\in\{e,i,s\}$, $\mu_z$ is tonic drive to population $z$, $g_{zw}$ is the coupling strength from population $z$ to population $w$, $\tau_{mz}$ is the membrane time constant, $\Delta_z$ is the degree of heterogeneity, $s_z$ is the synaptic time course of each population, $\tau_z$ is the synaptic decay time constant, $i_z(t)$ is time-dependent input, and $\zeta_{jz}$ are random numbers drawn from the Cauchy distribution with density function, $f(\zeta)=1/[\pi(1+\zeta^2)].$  the times, $t_{zj}$ are defined as
\[
\lim_{t\to t_{zj}}v_{zj}(t)=+\infty.
\]
When $v_{zj}(t^-)=\infty$, $s_z(t)$ is incremented by $\tau_{mz}/400\tau_z$ and $v_{zj}(t^+)=-\infty$.  
For the purposes of simulations, we make the transformation $v(t)\tan(\theta(t)/2)$ \cite{ke86} so that $\theta=\pi$ corresponds to $v(t)=+\infty$ and $\theta=-\pi$ corresponds to the reset to $-\infty$. Note that on the circle, $-\pi=\pi$ so that with this transformation all the behavior is continuous.  The equation:
\[
v'=a + b v^2
\]
becomes
\[
\theta' = b(1-\cos\theta)+a(1+\cos\theta).
\]
The spiking model is integrated using the Euler method with a time stem of 0.02 msec.

Phase relationships between the spikes and $s_e(t)$ are found by using a 1000 msec stretch at steady state and identifying the peaks of $s_e(t)$.  The phase of a spike is determined by finding the closest $s_e(t)$ peak ($t_{peak}^j$) that precedes the spike and then setting
\[
\theta=360 \frac{t_{spike}-t_{peak}^j}{t_{peak}^{j+1}-t_{peak}^j}.
\]

The mean-field model (see main text) is integrated using CVODE (a stiff integrator). All simulations and bifurcation diagrams are created using XPPAUT \cite{xppaut}. Codes are available from the authors.

\section*{Author Contributions}
\begin{description}
\item {\bf Conceptualization:} Bard Ermentrout
\item {\bf Data curation:} Arnab Dey Sarkar
\item {\bf Formal analysis:} Arnab Dey Sarkar, Bard Ermentrout
\item {\bf Funding acquisition:} Bard Ermentrout
\item {\bf Investigation:} Arnab Dey Sarkar  
\item {\bf Supervision:} Bard Ermentrout
\item {\bf Writing - original draft:} Arnab Dey Sarkar
\item {\bf Writing - review \& editing:} Bard Ermentrout  
\end{description}

\bibliography{newref}

\end{document}